\documentclass[aps,preprintnumbers,amsmath,amssymb,11pt]{revtex4}

\usepackage{epsf}

\usepackage{graphicx}

\usepackage{bm}

\begin {document}

%%%%%%%%%%%%%%%%%%%%%%%%%%%%%%%%%%%%%%%%%%%%%%%%%%%%%%%%%%%%%%%%%%%%%%%%%%%%%%%

%\def\Re{\operatorname{Re}}

\def\be{\begin{equation}}
\def\ee{\end{equation}}
\def\bea{\begin{eqnarray*}}
\def\eea{\end{eqnarray*}}

%%%%%%%%%%%%%%%%%%%%%%%%%%%%%%%%%%%%%%%%%%%%%%%%%%%%%%%%%%%%%%%%%%%%%%%%%%%%%%%

\title
    {
Chiral Quark Dynamics and Topological Charge: The Role of the Ramond-Ramond $U(1)$ Gauge Field in Holographic QCD
    }

\author {H. B. Thacker, Chi Xiong, and Ajinkya Kamat}
\affiliation
    {%
 Department of Physics,
    University of Virginia,
    P.O. Box 400714
    Charlottesville, VA 22901-4714\\
}

\date{\today}

\begin {abstract}%
The Witten-Sakai-Sugimoto construction of holographic QCD in terms of D4 color branes and D8 flavor branes
in type IIA string theory is used to investigate the role of topological charge in the chiral
dynamics of quarks in QCD. The QCD theta term arises from a compactified 5-dimensional Chern-Simons term on the
D4 branes. This term couples the QCD topological charge to the Ramond-Ramond $U(1)$ gauge field of
IIA string theory. The nonzero topological susceptibility of pure-glue QCD can be attributed to the presence of 
D6 branes, which constitute magnetic sources of the RR gauge field. The topological charge of QCD 
is required, by an anomaly inflow argument, to coincide in space-time with the intersection of the D6 branes and
the D4 color branes. This clarifies the relation between D6 branes and the coherent, codimension-one topological
charge membranes observed in QCD Monte Carlo calculations. Using open-string/closed-string duality, 
we interpret a quark loop (represented by a D4-D8 open string loop) in terms of closed-string exchange 
between color and flavor branes. The role of the RR gauge field in quark-antiquark annihilation processes is 
discussed. RR exchange in the s-channel generates a 4-quark contact term which produces an $\eta'$ mass insertion and provides an explanation for 
the observed spin-parity structure of the OZI rule. The $(\log {\rm Det\;U})^2$ form of the $U(1)$ anomaly emerges naturally. 
RR exchange in the t-channel of the $q\overline{q}$ scattering amplitude produces a Nambu-Jona Lasinio interaction which
may provide a mechanism for spontaneous breaking of $SU(N_f)\times SU(N_f)$. 
    {%

. }% 
\end {abstract}

\maketitle
\thispagestyle {empty}

 %%%%%%%%%%%%%%%%%%%%%%%%%%%%%%%%%%%%%%%%%%%%%%%%%%%%%%%%%%%%%%%%%%%%%%%%%%%%%%%

\section {Introduction}
\label{sec:intro}
In the holographic construction of QCD-like gauge theories from type IIA string theory \cite{Witten98,Sakai04}, the
elementary fields of the gauge theory are associated with the low-lying spectrum of open strings on $N_c$ coincident D4 branes and $N_f$ coincident D8 branes whose intersection is 
4-dimensional spacetime.. 
Gluons and quarks are represented by open D4-D4 and D4-D8 strings, respectively. In the closed string sector, the $U(1)$ gauge field associated with the massless
states of the Ramond-Ramond (RR) string plays a special role in reproducing the low energy chiral dynamics associated with topological charge in QCD \cite{Witten98}.
In the holographic theory, the QCD theta term arises from a 5-dimensional Chern-Simons term which couples the RR gauge potential to the topological charge of the color
gauge field. A very useful way of understanding the origin of this Chern-Simons term and its implications for low energy hadron physics 
is to interpret it in terms of anomaly inflow \cite{Callan-Harvey,Green}.
By standard arguments, the coupling of the RR field to the color gauge field on the D4 branes is dictated by the requirement that chiral anomalies on the brane, associated with the topological
charge of the gauge field, should be cancelled in the higher dimensional bulk theory by the inflow of Ramond-Ramond flux. Equivalently, in the presence of D-branes, 
the equations of motion and Bianchi identity for the RR field must be modified to include electric and magnetic source terms on the world volume of the branes. The form
of these source terms is dictated by the anomaly inflow argument, which determines the $U(1)$ gauge variation of the RR field that must accompany a Yang-Mills gauge 
transformation in order to define a gauge invariant RR field strength. In terms of the RR potential $C_1$, the gauge
invariant field strength is no longer just $dC_1$, but must include a Yang-Mills source term on the D4 branes proportional to the Chern-Simons current of the color gauge field.

In order to account for nonzero topological susceptibility, there must be a massless pole in the correlation function of two Chern-Simons currents with residue equal to $\chi_t$ \cite{Luscher78}.
This does not imply a massless physical particle, since the Chern-Simons current is not gauge invariant. On the other hand, this pole does have a direct physical significance, in that it
gives rise to a delta-function contact term in the gauge invariant topological charge correlator. Such a contact term is expected on general principles \cite{Seiler87}.
Since all real intermediate state contributions to $\chi_t$ (as computed from the integrated Euclidean correlator)
are negative, the correlator must be dominated by a positive contact term. The presence of such a contact term has been confirmed by
lattice calculations \cite{Horvath03,Horvath05b}. The Monte Carlo results showed that this contact term is the result of a laminated vacuum structure consisting of
interleaved, alternating-sign codimension one sheets of topological charge \cite{Horvath03}.  
From a holographic viewpoint, the massless pole required in the Chern-Simons current correlator to obtain finite topological susceptibility is associated with the exchange of 
a massless Ramond-Ramond gauge boson. The Ramond-Ramond field is a holographic description of the fluctuations of the topological charge membranes \cite{Thacker10}. 
The physical manifestation of RR gauge boson exchange in pure glue QCD at low energy is the appearance of a contact term in the gauge invariant 
topological charge corrrelator.  

\begin{figure}
\vspace*{4.0cm}
\includegraphics{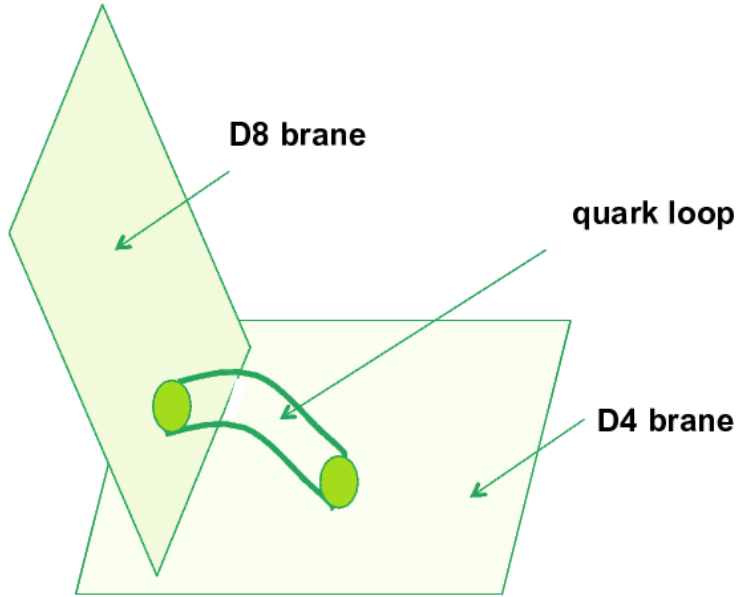}
\vspace{6.5cm}
\caption{An open string quark loop as a closed string exchange between D4 and D8 branes.}
\label{fig:D4D8string}
\end{figure}

We find other interesting phenomenological effects of RR exchange by considering open-string/closed-string duality (OS/CS) for quark loops.
The quark is represented by an open string attached at one end to a color D4 brane and at the other end to a flavor D8 brane. A closed quark loop in the field theory
is represented by a one-loop open string diagram which can be reinterpreted via OS/CS duality as the tree-level exchange of a closed Ramond-Ramond string 
between the flavor branes and the color branes (see Fig. \ref{fig:D4D8string}). 
This provides an interesting separation between the flavor and color structure of the quark loop. This closed-string description of
a quark loop is particularly appropriate for discussing its ultraviolet behavior, which is determined in the string theory by low-energy structure in the closed string channel.
Just as anomaly inflow arguments can be used to determine the coupling of the RR field to the color branes, similar arguments determine the RR coupling to the flavor branes
\cite{Sakai04}. In this case, anomaly inflow relates the RR gauge field to the $U(1)$ chiral Goldstone ($\eta'$) field, giving the familiar equivalence between a shift of $\theta$ and a chiral
rotation. By arguments similar to those leading to the contact term in the topological charge correlator, we show that RR exchange between quark lines induces an effective 
4-quark contact term. In the flavor singlet channel for quark-antiquark scattering, s-channel RR exchange provides the $\eta'$ mass insertion quark diagram (``double hairpin'' graph, Fig. \ref{fig:hairpin}).
In addition to reproducing the Witten-Veneziano relation for the $\eta'$ mass \cite{Witten79b,Veneziano79}, the RR exchange picture provides an explanation of the pure double pole behavior observed in
Monte Carlo calculations of the quenched double-hairpin correlator \cite{Bardeen04}. It has long been known that at large $N_c$, chiral Lagrangian arguments \cite{DiVecchia80,Witten_largeN,Rosenzweig80} lead 
to a specific form for the effective action term which correctly describes the axial $U(1)$ anomaly. This term must have the form of a pure $\eta'$ mass insertion:
\begin{equation}
\label{eq:U1breaking}
{\cal L}_a \propto - (\log{\rm Det} U-\log{\rm Det} U^{\dag})^2
\end{equation}
where $U_{ab}\propto \overline{q}_a(1+\gamma_5)q_b$ is the $N_f\times N_f$ chiral field.  This large $N_c$ expectation is quite distinct from the effective action induced by instantons, which 
would be proportional to the chiral determinant itself and therefore include OZI-violating six-quark interactions. We will show that an effective action
of the form (\ref{eq:U1breaking}) is exactly what is expected from the 4-quark contact term generated by RR exchange. This follows essentially from the fact that
the RR field couples, by anomaly inflow, to the $U(1)$ phase of the chiral field. Thus, the anomaly term (\ref{eq:U1breaking}) is interpreted as describing a contact term generated by RR exchange
between two chiral densities.

In addition to its role in QCD topological susceptibility and the $\eta'$ mass insertion, RR exchange may also play a direct role in spontaneous chiral symmetry
breaking and the formation of the chiral condensate. The $\eta'$ mass insertion arises from the $q$-$\overline{q}$ annihilation diagram (Figure \ref{fig:hairpin}), i.e.
s-channel RR exchange. This diagram appears only in the flavor singlet channel and explicitly breaks axial $U(1)$ symmetry. But there is also a 4-quark contact term generated
by RR exchange in the t-channel of the quark-antiquark amplitude (Figure \ref{fig:valence}). This interaction is the same for flavor singlet and nonsinglet channels. It is
a $U(N_f)\times U(N_f)$ conserving Nambu-Jona Lasinio-type interaction, suggesting that RR exchange may be the driving force behind $S\chi SB$.   
Finding that topological fluctuations play a role in the formation of the chiral condensate would not be that surprising. An analogous mechanism in the framework of the instanton 
liquid model has been discussed \cite{Diakonov84}. In that model, the 'tHooft near-zero modes of the instantons provide the eigenstates occupied by the quarks in the chiral condensate.
Similarly, in a vacuum of topological charge membranes (D6 branes), the chiral condensate forms from the surface modes of the quarks on the codimension one brane
surfaces. RR exchange in the t channel provides an attractive interaction between a D6 brane and a neighboring anti-D6 brane, and hence an attractive 
quark-antiquark interaction.

\section{Instantons and topological charge membranes as electric and magnetic sources of
Ramond-Ramond field}

\begin{figure}
\vspace*{4.0cm}
\includegraphics{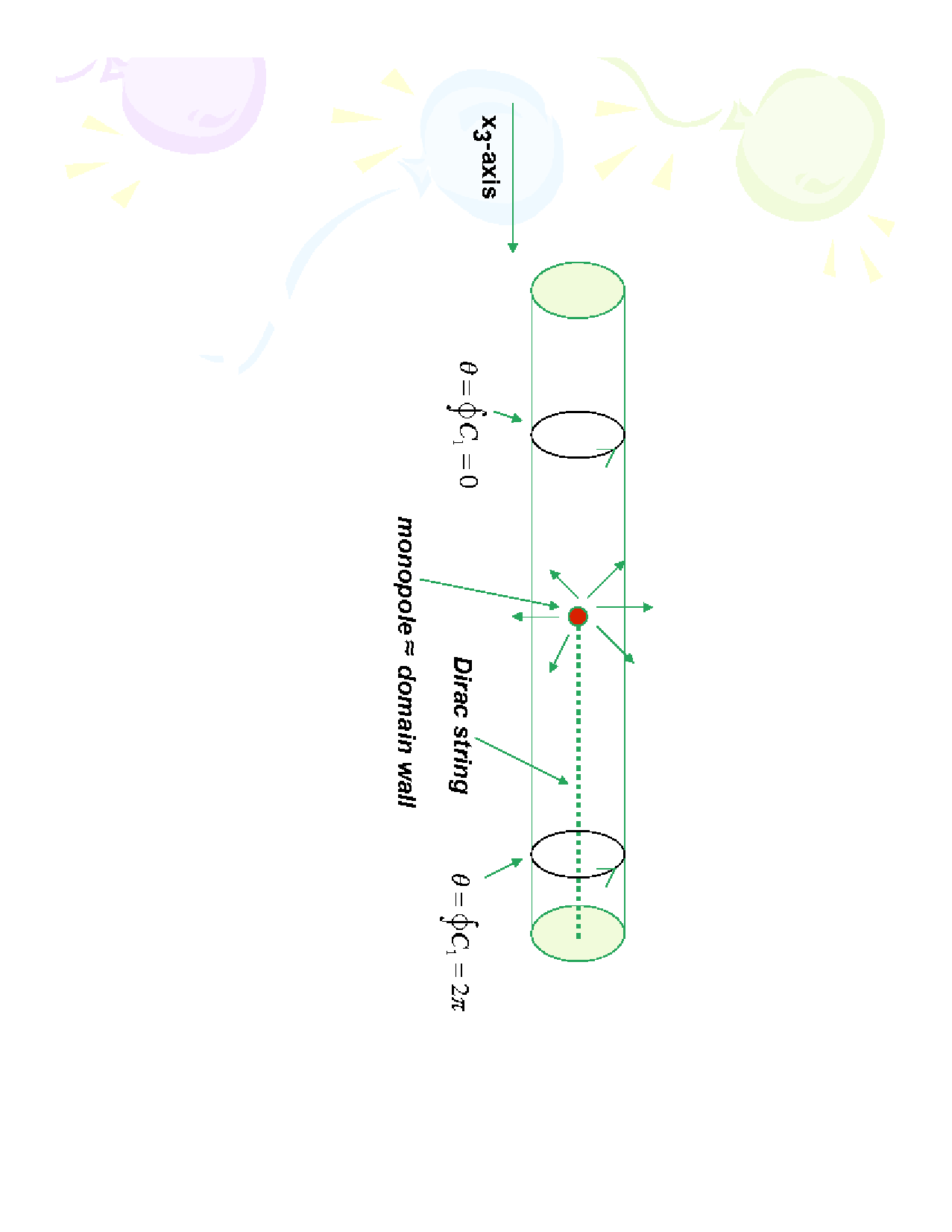}
\vspace{4.5cm}
\caption{View of a D6 brane as a magnetic monopole of the Ramond-Ramond gauge field in the 3 dimensions transverse to its worldvolume.}
\label{fig:laughlin}
\end{figure}

The origin of the QCD theta term was first discussed in the holographic framework in Ref. \cite{Witten98}. 
In Witten's construction, the color D4 branes are compactified around an $S_1$ with SUSY breaking boundary conditions.
The QCD theta term arises from a 5-dimensional Chern-Simons term which couples the Wilson line of the RR gauge 
potential $C_1$ to the topological charge density of the color D4 brane gauge field,
\begin{equation}
\label{eq:Chern-Simons}
S_{CS} = \int_{D4} C_1\wedge Tr\left(F\wedge F\right) \;\; .
\end{equation}
Here, $Tr$ is a trace over color indices. 
As shown in Ref. \cite{Green} and discussed below, the form of the
Chern-Simons term is dictated by anomaly inflow arguments.
In the full IIA string theory, this term represents the fact that fluctuations of the color gauge fields on the D4 brane
can absorb and emit closed RR string states which propagate in the bulk. In the field theory limit,
after compactification around the $S_1$ (with the circumference of the $S_1$ playing a role analogous to lattice spacing in lattice QCD), the surviving part of this Chern-Simons term is proportional to 
the RR Wilson line around the compact direction,
\begin{equation}
\label{eq:theta}
\int_{S_1} C_1 \equiv \theta(x)
\end{equation}
Here we denote 4-dimensional spacetime coordinates by $x$. If $\theta$ is a spacetime constant $\theta_0$, the CS
term reduces to a QCD theta term,
\begin{equation}
S_{CS}\rightarrow \theta_0\int_{R_4} Tr(F\wedge F)  \; .
\end{equation}
where $R_4$ is 4-dimensional spacetime.
As shown by Witten \cite{Witten98}, topological fluctuations in
the 4-dimensional gauge theory at large $N_c$ should be in the form of domain walls, i.e. codimension one membranes, which are described
by wrapped D6 branes in the string theory. This can be contrasted with an instanton liquid model, which would correspond to
D0 brane excitations. With respect to the 10-dimensional RR U(1) gauge field, D0 branes and D6 branes can be regarded as electric
and magnetic sources, respectively. From this point of view, the transition from the perturbative to the physical,
confining vacuum is caused by a condensation of magnetic Ramond-Ramond charge into alternating-sign layers \cite{Thacker10}. 
In the large-$N_c$ physical vacuum, populated by D6 branes, the role of the Chern-Simons term (\ref{eq:Chern-Simons}) is crucial
even when the overall QCD theta parameter is set to zero. In fact, in the presence of magnetic D6 brane sources, the RR potential $C_1$
cannot be uniquely specified globally, but must be defined in terms of overlapping sections, as with Dirac monopoles in 4D Maxwell theory.
Because of this nonuniquenes of $C_1$, it is more meaningful to write the Chern-Simons term in the form obtained by 
integrating by parts,
\begin{equation}
\label{eq:CS2}
S_{CS}\rightarrow -\int_{D4}dC_1\wedge {\cal K}
\end{equation}
where ${\cal K}$ is the 3-form satisfying $d{\cal K} = Tr(F\wedge F)$. Specifically, 
\begin{equation}
\label{eq:CS3form}
{\cal K} = Tr\left(A\wedge F-\frac{1}{3}A\wedge A\wedge A\right)
\end{equation}
Here and elsewhere the color gauge field potential on the D4 branes is denoted by $A$. The D8 brane gauge potential will be called ${\cal A}$.
We associate the 3-form (\ref{eq:CS3form}) with the 3-dimensional intersection between the D4 branes and the D6 brane. We will choose axes in 4D spacetime such that the 
D4-D6 ``I-brane'' intersection spans the $x_0$, $x_1$, and $x_2$ coordinates, with $x_3$ being the spacetime direction transverse to the I-brane (and hence
transverse to the topological charge membranes of the color gauge field). (From detailed Monte Carlo studies \cite{Horvath03,Horvath05,Ahmad05,Ilgenfritz}, the 
topological charge membranes are expected to be locally flat over a distance scale comparable to the confinemenet scale. Over larger distances,
the branes bend and fold and their orientation decorrelates.)
In the construction of Ref. \cite{Witten98}, the three dimensions transverse to the D6 brane consist of the holographic (radial) direction $x_5$, the compact $S_1$ (angular)
direction $x_4$, and one spacetime direction, which we take to be $x_3$. This 3D space can be represented
as a solid cylider, as depicted in Fig. \ref{fig:laughlin}. The D6 brane is a point magnetic RR charge in this space, and the construction of
a nonsingular gauge potential which is continuous away from the source requires $C_1$ to be defined separately on two disks or hemispheres on opposite
sides of the D6 brane and matched together by a topologically nontrivial $U(1)$ gauge transformation around the equator, leading to the quantization of
D6 brane charge. By Eq. (\ref{eq:theta}), this quantization enforces the fact that the value of $\theta(x)$ on opposite sides of the D6 brane must differ by 
an integer multiple of $2\pi$. So, rather than reducing to a constant theta parameter in the field theory limit, $\theta(x)/2\pi$ becomes an integer valued
field which jumps by $\pm 1$ at the location of a D6 brane. With the Chern-Simons action written in the form (\ref{eq:CS2}),
the codimension one topological charge membrane in the color gauge field appears
at the intersection of the D6 brane and the D4 branes, where low-mass open D4-D6 strings reside. 
Open-string/closed-string duality implies a coupling due to massless RR exchange between
D4 and D6 brane. Just as an instanton in the gauge theory can be interpreted as a D0 brane bound to the D4 color branes \cite{Witten96}, the topological charge membrane can be seen
as a D6 brane (wrapped on $S_4$) bound to the color branes. 

\section{Anomaly inflow and Ramond-Ramond couplings}

The Chern-Simons 3-form ${\cal K}$ in (\ref{eq:CS2}) is not invariant under a color gauge transformation. To maintain overall gauge invariance, we must
assume that a color gauge transformation is accompanied by a $U(1)$ transformation of the RR gauge field $C_1$. The RR field strength $G_2$ satisfies
an anomalous Bianchi identity, $G_2\neq dC_1$, which includes a magnetic source term arising from the color gauge field,
\begin{equation}
\label{eq:G2}
G_2 = dC_1 - \mu \left[*({\cal K}\wedge \delta_{D_4})\right]
\end{equation}
Here, $\delta_{D_4}$ is a 5-form given by a product of delta functions in the coordinates transverse to the D4 brane,
so the second term in (\ref{eq:G2}) is nonvanishing only on the D4 brane surface. In this term the $*$ represents the Hodge dual in 10-dimensional space
of the 8-form ${\cal K}\wedge\delta_{D_4}$. It is a 2-form with indices that are both on the D4 branes and transverse to the 0-1-2 I-brane. Thus the source term
from the D6 brane contributes only to the 3-4 component of the Ramond-Ramond field.
$\mu$ is fixed by the quantization condition. In terms of the D4 brane tension $T_4$ it is given by $\mu = T_4/16\pi^2$.
Under a Yang-Mills transformation $A\rightarrow g^{-1}Ag+g^{-1}dg$, the RR field transforms 
in a manner dictated by the descent equations \cite{Green}. This is most easily expressed in terms of the dual magnetic RR field 
strength $G_8=*(G_2)$ and associated potential $C_7$,
\begin{equation}
\label{eq:G8}
G_8 = dC_7 - \mu ({\cal K}\wedge \delta_{D_4})
\end{equation}
It is the potential $C_7$ that couples directly (i.e. minimally) to the 7-dimensional world volume of the magnetically
charged D6 branes. Under a Yang-Mills transformation $g$, this transforms as
\begin{equation}
\delta(dC_7) = \mu(\delta{\cal K}\wedge\delta_{D_4})
\end{equation}
The explicit transformation of the Chern-Simons 3-form is given by
\begin{equation}
\label{eq:gaugeK}
\delta{\cal K} = \mu\left[d\; Tr(dg\;g^{-1}\wedge A)\; + \;\frac{1}{3}Tr(g^{-1}dg\wedge g^{-1}dg\wedge g^{-1}dg)\right]
\end{equation}
The second term in (\ref{eq:gaugeK}) is proportional to the winding number density of the 3-dimensional gauge transformation on the I-brane,
\begin{equation}
w(x) = \frac{1}{24\pi^2}Tr(g^{-1}dg\wedge g^{-1}dg\wedge g^{-1}dg)
\end{equation}
With appropriate boundary conditions on $g$, $w(x)$ integrates
to an integer over the 3-dimensional I-brane. The quantization of this Chern-Simons winding number term incorporates the $2\pi$ step function discontinuity
in $\theta(x_3)$ associated with the presence of a D6 brane. A Yang-Mills transformation
$g$ with nonzero winding number on the 3-dimensional I-brane describes the matching of gauges on the two sides of the brane
needed to implement the anomalous Bianchi identity for the RR U(1) field.

The gauge invariance of the RR field strength $G_2$, Eq.(\ref{eq:G2}), ties the behavior of the bulk field $dC_1$ to fluctuations of the Wilson bag 
operator ${\cal K}$ on the D4 brane world volume. To see the implications of this connection for the 4-dimensional gauge theory, 
let us consider the D4 brane gauge theory in the presence of a D6 brane source. In the field theory limit, the relevant component of the RR field strength is
$\left(G_2\right)_{3 4}$ where $x_3$ is the direction in 4-dimensional spacetime transverse to the D6 brane, and $x_4$ is the compact $S_1$ direction.
Integrating (\ref{eq:G2}) around $S_1$, defining $\theta(x)$ by (\ref{eq:theta}), and writing indices explicitly, we have
\begin{equation}
\label{eq:intG2}
\oint_{S_1}\left(G_2\right)_{3 4} = \partial_3 \theta - K_3
\end{equation}
where we define the Chern-Simons current of the 4D Yang-Mills field ($\mu=0,1,2,3$)
\begin{equation}
K_{\mu} \equiv \epsilon_{\mu \nu \lambda \tau}Tr\left(A^{\nu}F^{\lambda\tau}-\frac{1}{3}A^{\nu}A^{\lambda}A^{\tau}\right)
\end{equation}
satisfying 
\begin{equation}
\partial^{\mu}K_{\mu}=\epsilon_{\mu\nu\sigma\tau}Tr(F^{\mu\nu}F^{\sigma\tau})
\end{equation}

The first term on the right hand side of (\ref{eq:intG2}) is a delta-function at the location of the D6 brane (because $\theta$ is a step function). 
A Yang-Mills gauge transformation induces a $U(1)$ gauge transformation of the second
term in (\ref{eq:intG2}) which must be cancelled by the variation of the first term. Roughly speaking, a gauge invariant excitation consists of the charged D6 brane and its attached color gauge 
field (a topological charge membrane), which must fluctuate together to form a physical propagating wave. The mechanism by which this generates a contact term in the topological
charge correlator and nonzero susceptibility is very similar to the ``Kogut-Susskind dipole'' mechanism in the 2-dimensional massive Schwinger model \cite{Kogut75}. It is useful to recall the
essential ingredients  of this simple example of anomaly inflow to point out the parallels with the D6-brane/topological-charge system. In the 2D massive Schwinger model one has
a choice of defining a conserved axial vector current $j^5_{\mu}$ which is not gauge invariant, or a gauge invariant current $\hat{j}^5_{\mu}$ which is not conserved:
\begin{eqnarray}
\partial^{\mu}j^5_{\mu} & = & 0 \\
\partial^{\mu}\hat{j}^5_{\mu} & = & \epsilon_{\mu \nu}F^{\mu \nu} 
\end{eqnarray}
The difference
between the two can be obtained by a point-splitting regularization, where the gauge invariant current acquires an extra term from the gauge link between the fermion operators.
This gives the correct form of the anomaly in 2D $U(1)$ gauge theory,
\begin{equation}
\label{eq:KSdipole}
\hat{j}^5_{\mu} = j^5_{\mu} - \epsilon_{\mu \nu}A^{\nu}
\end{equation}
In covariant Lorentz gauge quantization, the Hilbert space contains unphysical, negative metric states.
In the chiral limit, the terms on the right hand side of (\ref{eq:KSdipole}) can be represented in terms of 
a pair of massless scalar fields $\phi_1$ and $\phi_2$, where $\phi_2$ is a ghost field with negative norm.
Physical states only couple to the gauge invariant combination $\phi_1-\phi_2$. As a result, the massless poles
in the $\phi_1$ and $\phi_2$ propagators always cancel in physical amplitudes. The physical meson has a finite
mass due to the chiral anomaly (\ref{eq:KSdipole}). This is the Kogut-Susskind dipole mechanism.
For 4-dimensional QCD, the anomaly inflow constraint on the RR field strength, Eq. (\ref{eq:intG2}), plays a role analogous to (\ref{eq:KSdipole}) in the Schwinger model.
After we introduce quark flavors (see below), the quantity $\partial_{\mu}\theta$ will be identified with the flavor singlet axial vector current, so the first term on the right hand side of
(\ref{eq:intG2}) is the analog of the first term in (\ref{eq:KSdipole}) for the Schwinger model. The form of the second term, which represents the gauge  anomaly in these two equations, reflects
the structure of a codimension one domain wall in the two cases. In 2D the discontinuity is represented by a Wilson line, while in 4D Yang-Mills, the D6 brane
discontinuity is represented by a 3-dimensional Wilson bag integral \cite{Luscher78}.

By a generalization of the Kogut-Susskind pole cancellation mechanism, if we look at matrix elements of the gauge invariant field strength (\ref{eq:intG2}),
the pole due to the massless RR gauge boson is cancelled by a ``wrong sign'' pole in the correlator of Chern-Simons currents $K_{\mu}$,
\begin{equation}
\label{eq:RRprop}
\int d^4x e^{iqx}\langle (K_{\mu}(x) K_{\nu}(0)\rangle \stackrel{q\rightarrow 0}{\sim} \frac{q_{\mu}q_{\nu}}{(q^2)^2}\chi_t
\end{equation}
Since $\partial^{\mu}K_{\mu}$ is the gauge invariant topological charge, the topological susceptibility $\chi_t$ is given exactly by the residue of
the massless pole in the $K_{\mu}$ correlator (\ref{eq:RRprop}). By the anomaly inflow cancellation, this residue must be equal in magnitude to the residue of the RR gauge boson pole.
(Note that the ``wrong sign'' of the $K_{\mu}$ correlator that allows it to play the role of the Kogut-Susskind ghost is the same sign that
allows it to contribute a positive contact term to the otherwise negative topological charge correlator.)

\section{The Ramond-Ramond field and chiral quark dynamics}

So far we have only considered the anomaly inflow on the D4 color branes due to a D6 brane source. The inclusion of D8 flavor branes 
into the system \cite{Sakai04} allows us to consider the implications of the Ramond-Ramond gauge field for the chiral dynamics of quarks.
This is  most easily formulated in terms of the $U(N_f)\times U(N_f)$ chiral field constructed from quark bilinears,
\begin{equation}
U \propto \overline{q}(1+\gamma_5)q
\end{equation}
As discussed in Ref. \cite{Sakai04}, the chiral field is obtained from the Wilson line of the D8 brane gauge field taken along a U-shaped path on the 2-dimensional disk $D$ 
starting on the D8 brane and ending on the $\overline{\rm D8}$ brane at holographic infinity. Following Ref. \cite{Sakai04} we choose coordinates $y$ and $z$ on the disk which are, respectively,
transverse and parallel to the D8 brane. The chiral field is then give by the Wilson line of the D8 brane gauge field along the holographic direction,
\begin{equation}
U(x) = P\exp\left(-\int_{-\infty}^{\infty} dz' {\cal A}_z(x,z')\right)
\end{equation}
Of particular interest is the flavor singlet $\eta'$ meson field, which is given by the U(1) phase of the chiral field,
\begin{equation}
\eta'(x) = \frac{-if_{\pi}}{\sqrt{2N_f}}\log {\rm Det} U = \frac{if_{\pi}}{\sqrt{2N_f}}\int_{-\infty}^{\infty}dz' tr({\cal A}_z)
\end{equation}
(Throughout this paper, we use upper case $Tr$ to denote color traces and lower case $tr$ to denote a trace over flavor indices.)
We have seen that an anomaly inflow requirement on the color branes led to the result that only the combination of operators in (\ref{eq:G2}) is invariant under a color gauge
transformation. A similar anomaly inflow argument on the flavor-brane end of the quark string leads to the usual identification of the chiral $U(1)$ phase and the theta parameter,
which can be seen as follows:
The gauge invariant Ramond-Ramond field strength in the presence of D8 branes is given by \cite{Sakai04}
\begin{equation}
\label{eq:G2quark}
G_2 = dC_1 + i\; tr({\cal A})\wedge\delta_{D_8}\equiv dC_1 + i\;tr({\cal A})\wedge \delta(y) dy
\end{equation}
(The factor $tr({\cal A})$ here comes from the leading term in the expansion of the general Chern-Simons interaction of the form $dC_1\wedge {\rm tr}\left({\cal A}\wedge e^{i{\cal F}/2\pi}\right)$.)
Note that here a quark is an electric source of $G_2$, while in (\ref{eq:G2}), the topological charge membrane is a magnetic source.
If we consider the component $(G_2)_{45}$ and integrate it over the 2-dimensional disk, we get
\begin{equation}
\label{eq:G2etaprime}
\int_D (G_2)_{45} = \theta + \frac{\sqrt{2N_f}}{f_{\pi}} \eta'
\end{equation}
Thus the invariance of the RR field $G_2$ under a chiral gauge transformation on the D8 branes leads to the usual identification of $\theta$ with a chiral phase rotation. 
The Witten-Veneziano formula, relating the anomalous mass of the $\eta'$ meson to the topological susceptibility $\chi_t$ of pure glue
QCD is obtained in a straightforward way by the cancellation of massless poles required by anomaly inflow.
As we have seen, the gauge invariance of the RR field strength on the color branes required a cancellation between the massless RR gauge boson pole in the $dC_1$ correlator
and the massless pole in the $\langle K_{\mu}K_{\nu}\rangle$ correlator. The residue of the latter is the topological susceptibility $\chi_t$.
The expression (\ref{eq:G2etaprime}) for the gauge invariant RR field strength on the flavor branes imposes an anomaly inflow cancellation between the massless RR 
pole in $dC_1$ and the Goldstone pole in the $j_5^{\mu}$ correlator. Working to leading order in $1/N_c$, we assume $f_{\eta'} = f_{\pi}$ 
and write
\begin{equation}
\langle 0|\hat{j}^5_{\mu}|\eta'\rangle = f_{\pi} p_{\mu}
\end{equation}
so the residue of the Goldstone pole is $f_{\pi}^2m_{\eta'}^2$. Equating the two residues gives the Witten-Veneziano relation
\begin{equation}
 \chi_t = \frac{f_{\pi}^2m_{\eta'}^2}{4N_f}
\end{equation}
More directly, we may construct the effective action term that represents the effect of Ramond-Ramond exchange between quarks. 
The massless RR propagator in 4D momentum space has the same form as the Chern-Simons current correlator (c.f. Eq. (\ref{eq:RRprop}))
\begin{equation}
\tilde{G}_{\mu\nu} = -\frac{q_{\mu}q_{\nu}}{(q^2)^2}\chi_t
\end{equation}
In coordinate space, this satisfies
\begin{equation}
\partial^{\mu}\partial^{\nu}G_{\mu\nu}(x-y) = \chi_t \delta^4 (x-y)
\end{equation}
Combining the anomaly inflow constraints of (\ref{eq:intG2}) and (\ref{eq:G2etaprime}) we see that the RR gauge boson
couples to the $\eta'$ field with an effective action
\begin{equation}
\label{eq:eta2}
S_{int}=\frac{2N_f}{f_{\pi}^2} \int d^4x d^4y \partial^{\mu}\eta'(x) G_{\mu\nu}(x-y)\partial^{\nu}\eta'(y) 
=-\frac{2N_f\chi_t}{f_{\pi}^2}\int d^4x\; \eta'^2(x)
\end{equation}
Thus, the RR exchange model reproduces the chiral Lagrangian form of the axial anomaly that was originally deduced
from large $N_c$ and OZI rule arguments \cite{Rosenzweig80,DiVecchia80,Witten_largeN}, namely, a pure $\eta'$ mass term,
\begin{equation}
\label{eq:anomaly}
{\cal L}_{int} = \frac{\chi_t}{4} \left(\log{\rm Det} U -\log{\rm Det}U^{\dag}\right)^2
\end{equation}

\begin{figure}
\vspace*{4.0cm}
\includegraphics{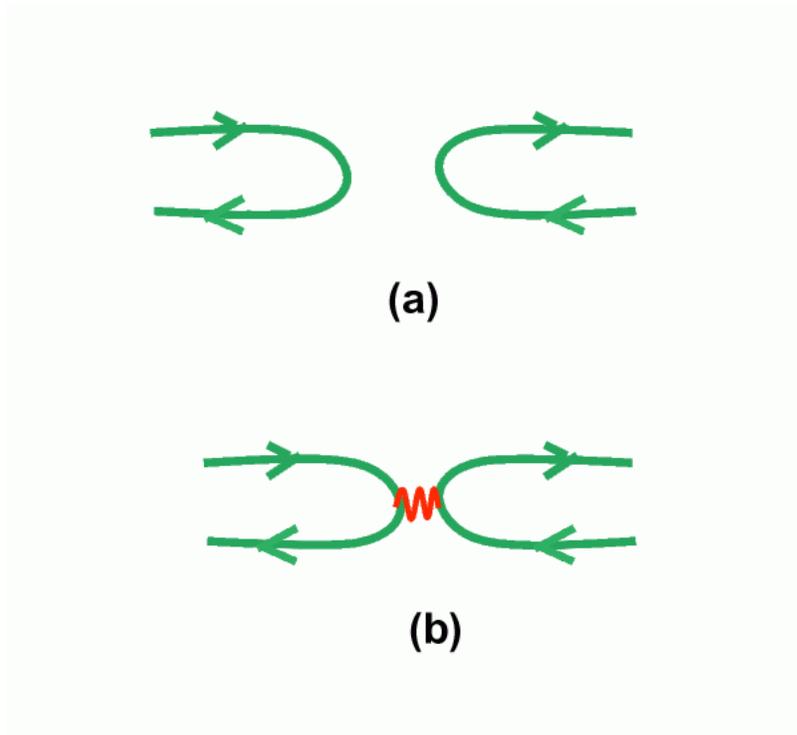}
\vspace{6.5cm}
\caption{(a) The quenched double hairpin correlator which measures the gluonic $\eta'$ mass insertion.
The indicated quark propagators are assumed to be summed over all gauge field configurations.
(b) The s-channel RR exchange picture for the hairpin correlator. Massless RR boson exchange results in a local
4-quark contact interaction due to its derivative coupling to the chiral field. These $q_{\mu}$ factors 
cancel the massless pole and convert it to a delta-function.}
\label{fig:hairpin}
\end{figure}

\begin{figure}
\vspace*{4.0cm}
\includegraphics{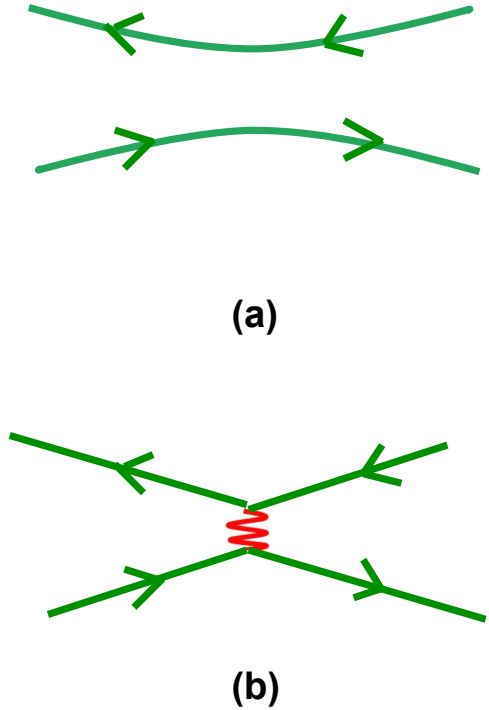}
\vspace{6.5cm}
\caption{(a) The valence diagram for quark-antiquark scattering in a meson. (b) The contribution of t-channel RR exchange to the valence diagram.}
\label{fig:valence}
\end{figure}

The RR exchange picture of the $\eta'$ mass insertion naturally incorporates a rather striking feature of the double
hairpin correlator, Fig. (\ref{fig:hairpin}),
\begin{equation}
G_{hp}(x) \equiv \langle \overline{q}_i\gamma_5 q_i(x)\;\overline{q}_j\gamma_5 q_j(y)\rangle \;\;\; i\neq j
\end{equation}
This correlator was studied in quenched QCD in Ref. \cite{Bardeen04}. and the time-dependence of the
zero-momentum correlator
\begin{equation}
\tilde{G}_{hp}(x_0) \equiv \int d^3 x G_{hp}(x)
\end{equation}
was measured in detail. Remarkably, over the entire range of time separations measured ($0<x_0<2\;fm$) 
this correlator is perfectly fit by a pure double pole formula \cite{Bardeen04},
\begin{equation}
\tilde{G}_{hp}(t) = const.\times (1+m_{\pi}t)e^{-m_{\pi}t}
\end{equation}
which is the zero 3-momentum fourier transform over $p_0$ of the quenched mass insertion diagram, Fig. \ref{fig:hairpin}, 
\begin{equation}
\frac{1}{p^2+m_{\pi}^2}m_0^2 \frac{1}{p^2+m_{\pi}^2}
\end{equation}
This is in marked contrast to the valence pion propagator, Fig. \ref{fig:valence}(a),
\begin{equation}
G_{val}(x) = \langle \overline{q}_i\gamma_5 q_j(x) \overline{q}_j\gamma_5 q_i(0)\rangle
\end{equation}
which approaches pure pion pole behavior at large separation, but exhibits substanitial excited state contributions at
shorter range. This implies that, in the hairpin correlator, all of the excited states created by the $\overline{q}\gamma_5 q$
operator (i.e. all states except the Goldstone boson) are ``filtered out'' by the hairpin vertex. 
Put another way, {\it quark-antiquark annihilation in flavor singlet pseudoscalar mesons takes place
only when the quark and antiquark are in a Goldstone boson state}. The RR gauge field couples to quark states via the chiral anomaly,
so it couples to the chiral condensate, which consists of surface quark zero modes on the Wilson bags/D6 branes. These are the states that
support the propagation of Goldstone bosons. Quark-antiquark pairs that are in excited, non-Goldstone pseudoscalar states
will therefore not couple to the RR field and not contribute to $q\overline{q}$ annihilation. 

Further information about the nature of quark-antiquark annihilation has been obtained from lattice studies of the spin-parity
structure of the OZI rule \cite{Isgur01}. It was found that, while the pseudoscalar hairpin correlator was easily measured and
gave a reasonably accurate determination of the $\eta'$ mass, the hairpin correlators in the vector and axial vector channels
(e.g. $\langle \overline{q}\gamma^{\mu}q(x)\;\overline{q}\gamma_{\mu}q(y)\rangle$)  are 
zero within errors, more than an order of magnitude smaller than the pseudoscalar case. This is also what is required by 
phenomenology, e.g. by the very small $\rho$-$\omega$ splitting compared to the large $\pi$-$\eta'$ splitting. In Ref \cite{Isgur01}
the scalar hairpin correlator was also found to be large and comparable in magnitude to the pseudoscalar one.
The s-channel RR exchange diagram, Fig. \ref{fig:hairpin}, provides a nice explanation for all of the observed properties of the
quark-antiquark annihilation process in QCD. 
Combining the absence of excited states in the pseudoscalar hairpin correlator with the vanishing of the hairpin correlator in
the vector and axial vector channels, we are led to conclude that quark-antiquark annihilation in QCD is limited to a very restricted
mechanism. To a good approximation, a quark and antiquark 
will annihilate only if they are in a state of zero total angular momentum, and, for the pseudoscalar channel, the
annihilating $q\overline{q}$ pair must be in the Goldstone ($\eta'$) state. Scalar $q\overline{q}$ pairs will also annihilate if both quark
and antiquark are in the condensate (i.e. they are occupying Dirac eigenmodes attached to the topological charge membranes). In the RR exchange
model of the quark-antiquark annihilation process, all of these properties follow from the fact
that the RR field couples directly to the chiral phase field $\propto \log{\rm Det}U$.

\section{Origin of the Nambu-Jona Lasinio 4-quark interaction}

We have seen that RR exchange in the gluon sector explains the existence of a positive contact term in the $Tr(F\wedge F)$ correlator and thereby,
the nonzero topological susceptibility of QCD. Extended to the quark sector, RR exchange in the s-channel of quark-antiquark scattering 
induces a 4-quark contact term which provides the annihilation vertex responsible for the $\eta'$ mass insertion. We next consider 
the effect of t-channel RR exchange in the quark-antiquark amplitude, Fig. \ref{fig:valence}(b). To understand this effect, let us rewrite the 
anomaly-induced $\eta'$ mass insertion as a 4-quark amplitude by expanding the log of the chiral field in (\ref{eq:anomaly}) in small 
fluctuations around its vacuum expectation value. Define
\begin{equation}
U_{ab} = \frac{1}{\langle \overline{q}q\rangle}\overline{q}_a(1+\gamma_5)q_b
\end{equation}
Then we can write $U=1+\delta U$ and expand to lowest order in $\delta U$,
\begin{equation}
\log{\rm Det}U=tr\log U\approx tr\delta U
\end{equation}
The $\eta'$ mass term can then be rewritten as a 4-quark interaction,
\begin{equation}
\label{eq:quarkanomaly}
\left(\log{\rm Det}U-\log{\rm Det}U^{\dag}\right)^2 \rightarrow \frac{1}{\langle \overline{q}q\rangle^2}\left(\sum_{a=1}^{N_f}\overline{q}_a\gamma_5q_a\right)^2
\end{equation}

\begin{figure}
\vspace*{4.0cm}
\includegraphics{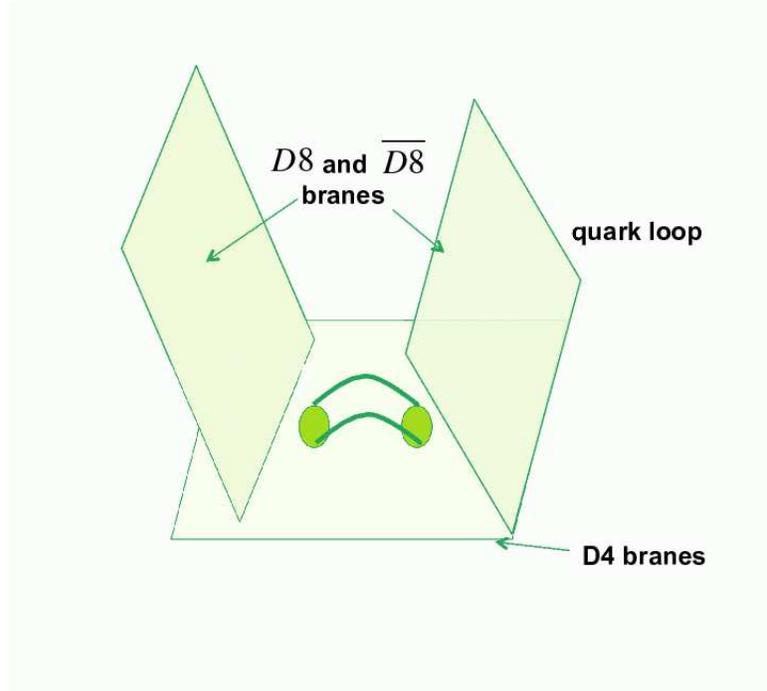}
\vspace{6.5cm}
\caption{String representation of the intermediate state in the double hairpin diagram for quark-antiquark scattering, which takes place by quark strings joining at the flavor ends.
Here the cylinder represents the partition function of the D4-D4 intermediate string.}
\label{fig:string-hairpin}
\end{figure}

\begin{figure}
\vspace*{4.0cm}
\includegraphics{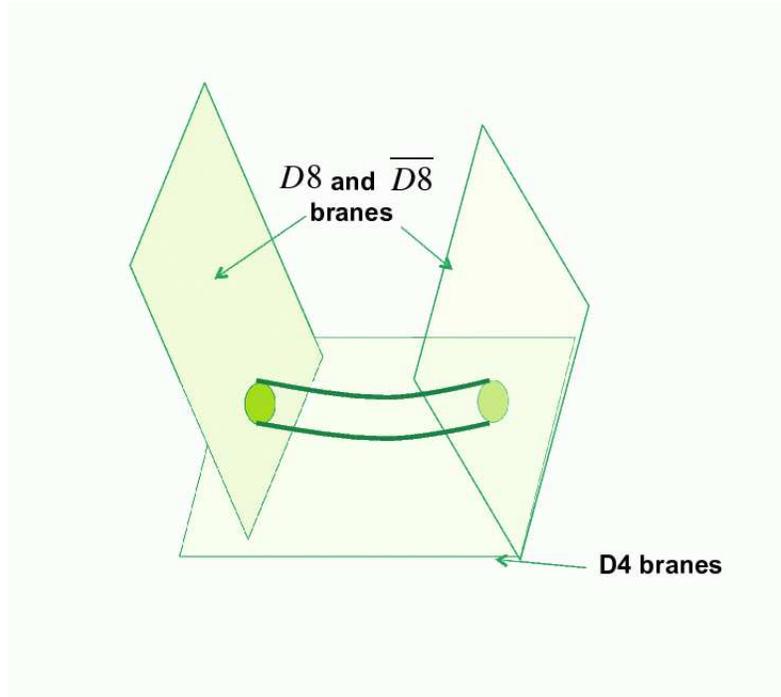}
\vspace{6.5cm}
\caption{String representation of the intermediate state in the valence diagram for quark-antiquark scattering, which takes place by quark strings joining at the color ends.
Here the cylinder represents the partition function of the D8-$\overline{\rm D8}$ intermediate string.}
\label{fig:string_valence}
\end{figure}

To determine the 4-quark interaction that is induced by t-channel RR exchange, we compare the string theoretic
view of quark-antiquark scattering to the field theory description. In a field-theoretic calculation, the pseudoscalar, flavor singlet meson propagator is given by the 
sum of valence (Fig. \ref{fig:valence}) and hairpin (Fig. \ref{fig:hairpin}) correlators. The nonsinglet Goldstone pion propagators are given by the valence
diagrams alone. Thus, the valence contribution to the correlator is invariant under $U(N_f)\times U(N_f)$ chiral transformations. The axial $U(1)$ anomaly comes entirely 
from the hairpin correlator. Now consider quark-antiquark scattering as the scattering of a D4-D8 and a D4-$\overline{\rm D8}$ string. The tree level open string 
scattering amplitude is given by the sum of two diagrams, corresponding to joining and splitting apart at either the color end or the flavor
end. If the flavor ends of the D4-D8 and D4-$\overline{\rm D8}$ strings join and separate, the intermediate state is a pure glue D4-D4 string, Fig. \ref{fig:string-hairpin},
so this corresponds to the hairpin annihilation diagram, Fig. \ref{fig:hairpin} in the field theoretic description. (In the Sakai-Sugimoto model, the joining of the flavor ends of the quark and
antiquark string is allowed by the fact that the D8 and $\overline{\rm D8}$ branes are joined together in the strong coupling region into a single U-shaped brane.)
The valence Feynman diagram, Fig. \ref{fig:valence}, corresponds to the scattering of quark strings by joining together at the color end, on the D4-brane world volume, Fig. \ref{fig:string_valence}.
This would be the only diagram for e. g. a charged pion, where the flavor ends of the two quark strings are on different D8 branes.
The intermediate state is a quark-antiquark D8-$\overline{\rm D8}$ string. Using open string/closed string duality, the D8-$\overline{\rm D8}$ string can be described
in the field theoretic limit by t-channel Ramond-Ramond exchange between quark and antiquark. We expect the Lorentz and flavor structure of the induced 4-quark
interaction to be similar to the 4-quark anomaly term, Eq. (\ref{eq:quarkanomaly}). However, because the flavor string-ends don't participate, this diagram does not depend on whether the
initial quark and antiquark were on the same or different flavor branes, i.e. whether the meson is a flavor singlet or nonsinglet. Thus, the effective 
interaction from t-channel RR exchange, Fig. \ref{fig:valence}(b) should be $U(N_f)\times U(N_f)$ invariant. Note that the $(\overline{q}\gamma_5q)^2$ anomaly term includes
both a $U(1)$ conserving term $2 q_L^{\dag}q_Rq_R^{\dag}q_L$ and a $U(1)$ violating term $(q_R^{\dag}q_L)^2 + h.c.$. 
By adding scalar $(\overline{q}q)^2$ interactions to cancel the $U(1)$ violating terms, we are led to an effective action for t-channel RR exchange of the form
\begin{equation}
\label{eq:t-channelRR}
{\cal L}_{t} \propto \left(\sum_{i=1}^{N_f} \overline{q}_i(1+\gamma_5)q_i\right) \left(\sum_{j=1}^{N_f}\overline{q}_j(1-\gamma_5)q_j\right)
\end{equation}
Here we defer a quantitative analysis to a subsequent paper and simply observe that the suggested form for t-channal RR exchange 
has the standard form of a $U(N_f)\times U(N_f)$ preserving Nambu-Jona Lasinio interaction \cite{NambuJona}. (A similar mechanism for generating an NJL interaction in holographic QCD
by integrating out the compactified component of the D4 brane gauge field has been discussed in \cite{Harvey06}.)
NJL models have an extensive and highly successful phenomenology.
Note that, although the interaction (\ref{eq:t-channelRR}) is pure flavor singlet in the t-channel, the Fierz transformed expression describes a $U(N_f)\times U(N_f)$ invariant 
interaction in the s-channel which is equal for flavor singlet and nonsinglet channels. In the usual NJL Bethe-Salpeter analysis, this will produce massless
pions in the nonsinglet channels.
It is a longstanding idea in QCD that topological charge fluctuations are responsible not only for resolving the $U(1)$ problem but also for driving the
spontaneous breaking of $SU(N_f)\times SU(N_f)$ and forming the chiral condensate \cite{Carlitz78,Diakonov84}. Previous discussions of topological charge driven $S\chi SB$ have
been framed in the context of the instanton liquid picture, in which the approximate 'tHooft zero modes of the instantons form the chiral condensate \cite{Diakonov84}.
From a slightly different viewpoint, Carlitz et al \cite{Carlitz78} argued that instantons would generate an attractive NJL quark-antiquark interaction
which could drive $S\chi SB$.
The RR exchange picture presented here suggests a reformulation of topological charge driven chiral symmetry breaking in the framework of holographic QCD.

\section{Discussion}

In many respects, the holographic framework for QCD provides an alternative to a lattice cutoff. The Monte Carlo evidence \cite{Horvath03,Ilgenfritz} that 
the QCD vacuum is dominated by a laminated array of topological charge membranes fits very naturally into the holographic framework \cite{Thacker10}. The lattice 
studies show clearly that the spacing between the alternating-sign membranes is fairly regular and of order a few lattice spacings, and remaining roughly constant
in lattice units for a range of correlation lengths \cite{Horvath03,Horvath05,Ahmad05}. This implies that the scaling limit of QCD follows a non-Landau-Ginsburg paradigm,
similar to that of antiferromagnetic systems, where alternating substructure at the lattice spacing scale can give rise to topological excitations in the scaling limit.
In the holographic formulation, the Ramond-Ramond gauge field describes the collective fluctuations of the layered arrangement of D6 and $\overline{D6}$ branes.
The Chern-Simons current $K_{\mu}$ of the color gauge field is related by anomaly inflow to $\partial_{\mu}\theta(x)$, i.e to the $\mu 4$ component of the
RR field strength. When crossing a D6 brane, $\theta(x)$ jumps by $2\pi$, representing the net outgoing RR flux from the charged D6 brane. By anomaly inflow, 
this discontinuity must coincide with the topological charge membrane of the color gauge field, represented in field theory by a ``Wilson bag''
integral of the Chern-Simons 3-form ${\cal K}$ over the 3-dimensional world volume of the bag surface. The $q^2=0$ pole in the $K_{\mu}$ correlator combines with the massless pole
in the RR correlator in a Kogut-Susskind dipole mechanism. In the K-S dipole for the 2D Schwinger model, the gauge invariant axial vector current operator describes the motion of a charged
fermion along with its attached gauge field. The massless pole in the conserved (non-gauge invariant) current matrix element is cancelled by the ghost pole 
describing the gauge field propagation. In holographic QCD the K-S dipole mechanism connects the motion of a charged D6 brane, represented by a $2\pi$ discontinuity
in $\theta(x)$, with the topological charge membrane excitation in the color gauge field, represented by the Chern-Simons operator ${\cal K}$.

It may seem that the appearance of the Ramond-Ramond field, in the form of a spacetime dependent $\theta(x)$, represents an additional degree of freedom in 
the holographic framework compared to, say lattice QCD, where $\theta$ is assumed to be a constant $\theta_0$ (usually zero). But in fact the spacetime
dependent $\theta(x)$, which is constant between D6 branes with a $\pm 2\pi$ discontinuity at the brane surfaces, can be interpreted as a singular gauge transformation which
smoothly connects the color fields on opposite sides of the brane. Note that the Wilson bag operator, given by the integral of ${\cal K}$ over a codimension one surface in spacetime 
is the gauge field operator which effectively inserts a discontinuity in $\theta$. (For example, a Wilson bag operator over a closed surface is equivalent to a theta term
in the interior of the bag, since $\partial^{\mu}K_{\mu}=Tr F\tilde{F}$.). Thus, in a purely 4-dimensional Yang-Mills framework the RR field $\theta(x)$ can be regarded as an
auxilary field which separates off the singular, sheet-like excitations in the Yang-Mills field and treats their dynamics separately. 
The requirement of overall gauge invariance of the RR field strength removes the redundancy of the auxiliary field and locks the motion of the $\theta(x)$ discontinuities to 
that of the topological charge membranes of the Yang-Mills field.
It is very clear from Monte Carlo studies \cite{Horvath03,Ahmad05}
that these singular, sheet-like gauge excitations are responsible for the positive contact term that appears in the Euclidean topological charge correlator. In this paper, we have shown that
such a contact term is the expected low-energy manifestation of Ramond-Ramond gauge boson exchange. This effect appears not only in the topological charge correlator, but also in the
anomaly-induced 4-quark contact term that produces an $\eta'$ ``hairpin'' mass insertion. We also discussed the possibility that RR exchange in the t-channel of quark-antiquark scattering 
generates a $U(N_f)\times U(N_f)$ symmetric Nambu-Jona Lasinio type interaction which could be responsible for the formation of the quark condensate. This provides an appealing physical
picture of spontaneous $SU(N_f)\times SU(N_f)$ breaking and Goldstone boson propagation and its relation to vacuum topological charge structure in gauge theory.  
The picture that emerges is of a left handed chiral condensate of $\overline{q}(1-\gamma_5)q$ living on the surfaces of the D6 branes interleaved
with sheets of right-handed $\overline{q}(1+\gamma_5)q$ on the surfaces of the anti-D6 branes. This is reminiscent of the instanton liquid model, in which left
and right condensate modes lived on instantons and antiinstantons, respectively, averageing out to a $\langle\overline{q}q\rangle\neq 0$ condensate. But a crucial difference
for the D6 brane condensate is that the quarks modes in the condensate are delocalized along codimension one sheets instead of being 'tHooft zero modes
confined to localized lumps. This provides a much more natural mechanism for Goldstone boson propagation than the mode-mixing or ``hopping'' that must be invoked in
instanton liquid models.

This work was supported by the Department of Energy under grant DE-FG02-97ER41027.

\end {document}